\documentclass[12pt]{article}
\pdfoutput=1
\usepackage{graphics, color}
\usepackage{graphicx}
\usepackage{amsmath}
\usepackage{amssymb}
\usepackage{amsfonts}
\usepackage[usenames,dvipsnames]{xcolor}
\usepackage[normalem]{ulem}
\usepackage[bottom]{footmisc}

\definecolor{orange}{rgb}{1,0.5,0}
\definecolor{grey}{rgb}{.5,.5,.5}
\definecolor{bluegreen}{rgb}{0,.5,.5}
\definecolor{darkgreen}{rgb}{0,.5,0}

\def\gsim{\, \rlap{$>$}{\lower 1.1ex\hbox{$\sim$}}\,}
\def\lsim{\, \rlap{$<$}{\lower 1.1ex\hbox{$\sim$}}\,}
\newcommand{\be}{\begin{equation}}
\newcommand{\ee}{\end{equation}}

\begin{document}

%Title page

\begin{titlepage}
\bigskip
\bigskip\bigskip\bigskip
%\centerline{\Large \bf Something something something}
\centerline{\Large \bf The Field Theory of Intersecting D3-branes}

\bigskip\bigskip\bigskip
\bigskip\bigskip\bigskip

 \centerline{{\bf Eric Mintun,}\footnote{\tt mintun@physics.ucsb.edu}*
 {\bf Joseph Polchinski,}\footnote{\tt joep@kitp.ucsb.edu }*${}^\dagger$ and {\bf Sichun Sun}\footnote{\tt sichun@uw.edu }${}^\dagger{}^\ddagger$
}
 \bigskip
\centerline{\em *Department of Physics}
\centerline{\em University of California}
\centerline{\em Santa Barbara, CA 93106 USA}
\bigskip
\centerline{\em ${}^\dagger$Kavli Institute for Theoretical Physics}
\centerline{\em University of California}
\centerline{\em Santa Barbara, CA 93106-4030 USA}
\bigskip
\centerline{\em ${}^\ddagger$Institute for Nuclear Theory}
\centerline{\em University of Washington Box 351550}
\centerline{\em Seattle, WA 98195-1550, USA}

\bigskip\bigskip\bigskip
%ABSTRACT

\begin{abstract}
We examine the defect gauge theory on two perpendicular D3-branes with a 1+1 dimensional intersection, consisting of $U(1)$ fields on the D3-branes and charged hypermultiplets on the intersection.  We argue that this gauge theory must have a magnetically charged soliton corresponding to the D-string stretched between the branes.  We show that the hypermultiplets actually source magnetic as well as electric fields.  The magnetic charges are confined if the hypermultiplet action is canonical, but considerations of periodicity of the hypermultiplet space in string theory imply a nontrivial Gibbons-Hawking metric, and we  show that there is then the expected magnetic kink solution.  The hypermultiplet metric has a singularity, which we argue must be resolved by embedding in the full string theory.  Another interesting feature is that the classical field equations have logarithmic divergences at the intersection, which lead to a classical renormalization group flow in the action.

\end{abstract}
\end{titlepage}

\baselineskip = 16pt
\tableofcontents

\baselineskip = 16pt

\setcounter{footnote}{0}

\section{Introduction}

Intersecting D-branes have many applications.  For example, they play a large role in top-down constructions of holographic duals.  It is therefore surprising to find that the low energy field theory on D3-branes intersecting in 1+1 dimensions has several interesting features that seem not to have been previously discussed.  In this paper, driven by an argument that this field theory must have magnetically charged solitons, we determine the low energy Lagrangian and discuss its properties. 

To begin, consider a pair of parallel D3-branes.  An F-string stretched between them is a BPS state which sources electric flux on the D3-branes.  The D3-brane system is invariant under $S$-duality, which takes the F-string to a BPS D-string which sources magnetic flux.  In the limit that the D3-branes are very close, the stretched F- and D-strings are much lighter than the string scale, and we should be able to describe them in an effective field theory.  As is well-known, the F-strings become the off-diagonal components of the ${\cal N}=4$ $U(2)$ fields, while the D-strings become magnetic monopole solitons in the spontaneously broken $U(2)$ theory.

Another possibility would have been for the low energy theory to contain independent magnetically charged fields.  However, these would produce large nonperturbative effects due to their $1/g$ couplings, and this is not consistent.  Such light nonperturbative states do arise in singular limits such as the conifold~\cite{Strominger:1995cz}, but for coincident D-branes perturbation theory should be valid. 

\begin{table}[h]
\begin{center}
\begin{tabular}{l|cccccccccc}
& 0 & 1 & 2 & 3 & 4 & 5 & 6 & 7 & 8 & 9 \\ \hline
D3 & x & x & & & x & x & & & & \\
D3$^\prime$ & x & x & & & & & x & x & &\\
F/D1 & x &&&&&&&&x&
\end{tabular}
\end{center}
\caption{The directions in which the two D3-branes and the stretched strings are extended.  An x signifies extension in the indicated direction.}
\label{braneTable}
\end{table}
Now consider two D3-branes oriented as in Table~1.  The only common spatial direction of the D3-branes is $x^1$, and we take them 
to be separated in the 8-direction.  We can now ask the same question: in the limit of small separation, how does the stretched D-string, which is a BPS state, appear in the  low energy effective theory?\footnote{This issue has been raised in Ref.~\cite{Constable:2002xt}.  Related questions have been discussed in Ref.~\cite{Gukov:2006jk}.} This effective theory has only  Abelian fields on the D3-branes, with charged hypermultiplets moving on the intersection, so it is not clear how a magnetically charged soliton could arise.  But independent magnetic degrees of freedom are excluded by the same argument as before.

In this paper we resolve this puzzle.  This will involve several novel properties of the intersecting brane theory.  In \S2 we study the effective low energy theory assuming canonical kinetic terms for the 3-3$'$ hypermultiplets, extending results of Ref.~\cite {Constable:2002xt}.  We show the 3-3$'$ fields are actually charged both electrically and magnetically under the D3 and D3$'$ gauge fields.  While this creates the possibility of magnetic solitons, we show that the magnetic charges are actually confined, unlike the D-strings that we are trying to find.  

In \S3 we show that the interpretation of the hypermultiplets in terms of brane reconnection requires the $U(1)$ D-terms to be periodic variables, and this implies a nonlinear kinetic term.  The hypermultiplet moduli space is a singular Gibbons-Hawking metric, with an infinite array of sources periodically identified.  This periodicity produces a sine-Gordon-like potential, which supports magnetically charged kink solitons.  We tentatively interpret the singularity as a breakdown of the low energy effective field theory of the intersection in regions of large field.

An interesting feature of systems with codimension-two defects, as here, is the presence of logarithmic divergences in the classical field theory.  Goldberger and Wise~\cite{Goldberger:2001tn}  showed that these could be treated by ordinary renormalization theory.  In \S4 we obtain the beta functions for the K\"ahler potential and for the Gibbons-Hawking harmonic function $V$.  The former is proportional to the squares of the moment maps, while the latter is simply a constant.  This renormalization mitigates, but does not remove, the singularity of the metric.  We show that the same effective action is obtained from the DBI picture of the intersection.

In \S5 we obtain the BPS equations, and argue that they have the desired soliton solutions.  We cannot solve them analytically, but we study them in various approximations.  Sec.~6 is discussion and conclusion.

\begin{table}[h]
\begin{center}
\begin{tabular}{l|cccccccccc}
& 0 & 1 & 2 & 3 & 4 & 5 & 6 & 7 & 8 & 9 \\ \hline
D3 & x & x &x & &  x&  & & & & \\
D5 & x & x & x& & &x & x & x & &\\
D3$''$ & x &&x&&&x&&&x&
\end{tabular}
\end{center}
\caption{A $T$-dual D3-D5 system}
\end{table}

The system we consider arises in various $T$-dual forms: we can dualize in one or more of the DD 2-, 3-, and 9- directions to produce higher dimensional versions, where the soliton is independent of the additional coordinates.  The most well-studied intersecting brane system is probably D3-D5 intersecting in 2+1 dimensions~\cite{Karch:2001cw}.  We can reach this by $T$-dualities in the 2- and 5-directions, reaching the configuration shown in Table~2.  However, this requires a $T$-dual in one of the ND-directions, and this changes the nature of the system.  In particular the would-be magnetic object is not part of the brane intersection: the D1 becomes a D3$''$ extended in the 5-direction, so it cannot end on the D3.  The various phenomena that we encounter are not present.

We were initially interested in this system as a simple field-theoretic model of $S$-duality, which it would inherit from the full string theory.  Since the gauge fields are Abelian, the hope was that the $S$-duality would take a rather simple form.  However, it is not clear that the brane system can be defined consistently as a field theory.  The singularities of the moduli space metric apparently require embedding in the full brane DBI theory, and this in turn can only be quantized by embedding in string theory.  If this is true, then there is no smaller $S$-dual system.  Nevertheless, our work shows that intersecting D-brane systems harbor several surprising properties: the magnetic couplings of the intersection fields, the nonlinear field space, and the classical renormalization.   These may have a variety of applications.

\section{Canonical action}
\label{MinimalGaugeSection}

The use of $d=4$, ${\cal N}=1$ superfields to write higher-dimensional theories has been developed in Refs.~\cite{Marcus:1983wb,ArkaniHamed:2001tb} and applied to intersecting branes in Ref.~\cite{Constable:2002xt}.  We first give the superfield form of the action, following most closely the presentation in~\cite{ArkaniHamed:2001tb}.  We then give the component form, demonstrating the magnetic couplings of the intersection fields.

\subsection{Symmetries and fields}

The brane system of Table \ref{braneTable} preserves eight supercharges.  The bosonic symmetries are
\be
SO(1,1)_{01}\times SO(2)_{45} \times SO(2)_{67} \times SO(4)_{2389} \times U(1)_V \times U(1)_{V'} \,. \label{sym}
\ee
On each D3-brane there lives the usual field content for a U(1) $d=4$, $\mathcal{N}=4$ gauge theory, but the supersymmetry algebras of the two branes are not the same.  In order to exhibit the common supersymmetry, it is useful to label the fields in a nonstandard way.  Consider the system obtained by $T$-duality in the 23 directions.  The D3-D3$'$ become D5-D5$'$.  Both D5 branes are extended in the 0123 directions, and the unbroken supersymmetry includes a $d=4$, $\mathcal{N}=1$ algebra in these directions.  We can write the theory in terms of superfields for this algebra.  The additional coordinates, $x^{4,5}$ for the D5 and $x^{6,7}$ for the D5$'$, are treated as parameters in the superfields.  This follows the strategy introduced in Refs.~\cite{Marcus:1983wb,ArkaniHamed:2001tb} for writing $d=10$, $\mathcal{N}=1$ SYM and other higher dimensional theories in terms of $d=4$, $\mathcal{N}=1$ superfields.  We then  dimensionally reduce in the 23 directions to obtain the system of interest.  The $T$-dual system has different global symmetries, but the fact that the dimensionally reduced system will have an $SO(4)_{2389}$ symmetry guarantees that it has $d=4$, $\mathcal{N}=2$ supersymmetry.

On the D5 brane, we will call the $d=4$, $\mathcal{N}=1$ vector multiplet $V$ and the scalar multiplets $Q_{1,2,3}$.  After dimensional reduction to the D3, 
the four component gauge field will be made up of the $A_{V0,1}$ components of the vector multiplet and the real and imaginary components of $Q_1$, while $A_{V2,3}$ will combine to become one of the complex SYM scalars.  The D5$^\prime$ brane is identical, with vector multiplet ${V'}$ and scalar multiplets $S_{1,2,3}$.  In this case, the components of $S_2$ will combine with $A_{{V'}0,1}$ to become the usual gauge field and $A_{V'2,3}$ becomes a complex scalar.  The scalars $A_{V2,3}$ and $A_{V'2,3}$ combine with the scalars $Q_3$ and $S_3$ respectively to become $SO(4)$ vectors, since these fields will describe the transverse coordinates of the branes in the $2389$ directions.  

On the intersection between the branes, there are two chiral fields $B$ and $C$, associated with the two orientations of F-strings stretching between the branes.  $B$ has fundamental charge $(1,-1)$ under the gauge groups $U(1)_V$ and $U(1)_{{V'}}$, while $C$ has charge $(-1,1)$.
 %Likewise, the fields $Q_1$ and $Q_2$ each transform as gauge fields under a different $SO(2)$.  
These are scalars under  $SO(4)$ and $SO(1,1)$.  Since the $SO(4)$ is associated with the $\mathcal{N}=2$ supersymmetry, this means that the scalars $B$ and $C$ are not transformed into each other via the supersymmetric R-symmetry.  Instead, it is fermionic superpartners of these fields that combine to form an $SU(2)_R$ doublet.  This differs from the construction in \cite{Seiberg:1994rs}, where the hypermultiplet scalars form an $SU(2)_R$ doublet.  Under $SO(2)_{45} \times SO(2)_{67}$,  $B$ and $C$ have charges 
$(\frac12, \frac 12)$.  They transform as spinors because the ND boundary conditions for the 4567 directions on the string means that the corresponding RNS fermions are periodic in the NS sector and generate spinorial states.

We will discuss the transformations of the superfields further in \S5, when we find the BPS equations.
%{\color{red} (Include discussion about the transformation of the supercharges?)}
\subsection{Superfield action}

We mostly follow the conventions of Wess and Bagger~\cite{Wess:1992cp} and the presentation of Ref.~\cite{ArkaniHamed:2001tb}.  In component form in Wess-Zumino gauge, we write
\be
B = B(y) + \sqrt{2} \theta \psi_{B} (y) + \theta \theta F_{B} (y)
\ee
using the same symbol for the scalar component as for the superfield itself.  It will be convenient to collect the spatial coordinates $x_4$ through $x_9$ into three complex coordinates $z_{123}$ with
\be
z_1 = \frac{1}{2} \left ( x_4 + i x_5 \right ) \quad \quad z_2 = \frac{1}{2} \left ( x_6 + i x_7 \right ) \quad \quad z_3 = \frac{1}{2} \left ( x_8 + i x_9 \right )
\label{znorm}
\ee
with derivatives denoted as $\partial_{z_1} = \partial_4 - i \partial_5$, etc.  Greek indices $\mu$, $\nu$, etc. will run over 0123 unless otherwise specified.  
  
The action for the 3-3 fields is then~\cite{Marcus:1983wb,ArkaniHamed:2001tb}
\be
\begin{split}
S_{\mbox{\scriptsize 3-3}} = \frac{1}{g_{\rm YM}^2} &\int d^2 x\, d x_4 dx_5 \Bigg \{   \int d^2 \theta\left [ \frac{1}{4} W^\alpha_V W_{V\alpha} + \frac{1}{2} \left ( Q_3 \partial_{z_1} Q_2 - Q_2 \partial_{z_1} Q_3 \right ) \right ]+ {\rm c.c.} \\  + &\int d^4 \theta \left [ \left ( \sqrt{2} \bar{\partial}_{z_1} V -  \bar{Q}_1 \right ) \left ( \sqrt{2} \partial_{z_1} V - Q_1 \right) - \bar{\partial}_{z_1} V \partial_{z_1} V + \bar{Q}_2 Q_2 + \bar{Q}_3 Q_3 \right ]  \Bigg \}\,.
\end{split} \label{33act}
\ee
As noted above, the $V, Q_i$ are functions of the parameters $x^{4,5}$, so they represent an infinite number of $d=4$, ${\cal N} = 1$ superfields.  We have dimensionally reduced in $x^{2,3}$, so only the integrations over the 0145 directions remain.  The terms involving ${z_1}$ derivatives of $V$ and $Q_1$ are fixed by the gauge invariance\footnote{Ref.~\cite{Constable:2002xt} gives the action in a slightly different form, which does not seem to be gauge invariant.  Incidentally, our notation here follows~\cite{ArkaniHamed:2001tb}, whose gauge parameter $\Lambda$ is   $i \Lambda_{\rm Wess\&Bagger}$.  %{\color{red} Need to check later equations.} 
}
\be
\begin{split}
V & \to V + \Lambda + \bar{\Lambda}
\\  Q_1 & \to Q_1 + \sqrt{2} \partial_{z_1} \Lambda \,,
\end{split}
\ee
where the superfield gauge parameter $\Lambda$ is also a function of $x^{4,5}$.  This choice of gauge transformation for $Q_1$ is consistent with its role as a gauge field,
\be
Q_1 = \frac{1}{\sqrt{2}} \left ( A_{V5} + i A_{V4} \right ) \,.
\ee
The terms involving ${z_1}$ derivatives of $Q_{2,3}$ are needed to give $SO(3,1)$ invariance in the 0145 directions for this part of the action.  Though both the $\mathcal{N}=2$ supersymmetry and the Lorentz invariance in the 45 directions are obscured in this form, in component form the relationship is much clearer.  This will be discussed in Section \ref{ComponentGaugeSection}.  Similarly the action for the $3'$-$3'$ fields is
\be
\begin{split}
S_{\mbox{\scriptsize 3$'$-3$'$}}  = \frac{1}{g_{\rm YM}^2} &\int d^2 x\, d x_6 dx_7 \Bigg \{   \int d^2 \theta\left [ \frac{1}{4} W^\alpha_{V'} W^{\vphantom\alpha}_{{V'}\alpha} + \frac{1}{2} \left ( S_1 \partial_{z_2} S_3 - S_3 \partial_{z_2} S_1 \right ) \right ] + {\rm c.c.} \\
  + &\int d^4 \theta \left [ \left ( \sqrt{2} \bar{\partial}_{z_2} {V'} -  \bar{S}_2 \right ) \left ( \sqrt{2} \partial_{z_2} {V'} - S_2 \right) - \bar{\partial}_{z_2} {V'} \partial_{z_2} {V'} + \bar{S}_1 S_1 + \bar{S}_3 S_3 \right ]  \Bigg \} \,. \label{3p3pact}
\end{split}
\ee

The fields $B$ and $C$ on the intersection are have charges $(1,-1)$ and $(-1,1)$ under $(V,V')$ respectively, so their  gauge transformations are
\be
\begin{split}
B & \to B e^{\Lambda' - \Lambda} \,,
\\ C & \to C e^{\Lambda - \Lambda^\prime} \,.
\end{split}
\ee
Here the superfield gauge parameters $\Lambda$ and $\Lambda'$ are evaluated at the defect $x_{4567}=0$.  The only gauge invariant combinations of $B$ and $C$ are then $|B|^2 e^{{V} - V'}$, $|C|^2 e^{{V'}-V}$, and $BC$. For a canonical kinetic term, these fields have the action
\be
\begin{split}
S_{\mbox{\scriptsize 3-3$'$}} = \frac{1}{g_{\rm YM}^2} \int d^4 x \Bigg \{  \int d^4 \theta \left ( |B|^2 e^{V-{V'}}  +  |C|^2 e^{{V'}-V}\right )  +  \frac{i}{\sqrt 2} \int d^2 \theta  \left ( B C Q_3 -  B C S_3 \right ) +{\rm c.c.} \Bigg \} \,, \label{33pact}
\end{split}
\ee
in which all fields living on the branes are evaluated at the defect.  The superpotential 
\be
W = \frac{i}{\sqrt2g_{\rm YM}^2}  \left ( B C Q_3 -  B C S_3 \right )
\ee
is fixed by $\mathcal{N}=2$ supersymmetry, since $V$ and $Q_3$ combine to form a single $\mathcal{N}=2$ vector multiplet.  Equivalently, this form is fixed by the requirement that we maintain the SO(4) symmetry of the original brane construction: $A_{V2,3}$ and the real components of the scalar $Q_3$ arise from the directions transverse to both branes, so $B$ and $C$ must couple identically to both.

To summarize, the full action~\cite{Constable:2002xt} is the sum of Eqs.~(\ref{33act}, \ref{3p3pact}, \ref{33pact}).
 
\subsection{Component action}
\label{ComponentGaugeSection}
 
Integrating out all the auxiliary fields gives the bosonic sector of the component form action,
\be
\begin{split}
S  = \frac{1}{g_{\rm YM}^2} \int_{\rm D3}\!\! d^2 x\, d x_4 dx_5 \Bigg \{&  - \frac{1}{4} F^{\mu \nu}_V F_{V \mu \nu} - \frac{1}{4} \left ( \bar{\partial}_{z_1} Q_1 + \partial_{z_1} \bar{Q}_1- \frac{1}{\sqrt{2}} \left (|C|^2 - |B|^2  \right ) \delta^2(x_4,x_5)\right )^2  \\ 
&\quad  -  \left | \partial^\mu Q_1 - \frac{i}{\sqrt{2}} \partial_{z_1} A_V^\mu \right|^2    - \partial_\mu \bar{Q}_2 \partial^\mu Q_2  - \partial_\mu \bar{Q}_3 \partial^\mu Q_3  - \partial_1 Q_3 \bar{\partial}^1 \bar{Q}_3 \\ 
& \quad- \left (\bar{\partial}_{z_1} \bar{Q}_2 - \frac{i}{\sqrt 2} \bar{B}\bar{C} \delta^2(x_4,x_5) \right ) \left ( \partial_{z_1} Q_2 + \frac{i}{\sqrt 2} BC \delta^2(x_4,x_5) \right ) \Bigg \} \\
+  \frac{1}{g_{\rm YM}^2} \int_{\rm D3'}\!\!\!  d^2 x\, d x_6 dx_7 \Bigg \{&  - \frac{1}{4} F^{\mu \nu}_{{V'}}F_{{V'} \mu \nu} - \frac{1}{4} \left ( \bar{\partial}_{z_2} S_2 + \partial_{z_2} \bar{S}_2+ \frac{1}{\sqrt{2}} \left (|C|^2 - |B|^2  \right ) \delta^2(x_6,x_7)\right )^2  \\ 
&\quad  -  \left | \partial^\mu S_2 - \frac{i}{\sqrt{2}} \partial_{z_2} A_{{V'}}^\mu \right|^2   - \partial_{z_2} S_3 \bar{\partial}_{z_2} \bar{S}_3 - \partial_\mu \bar{S}_3 \partial^\mu S_3   - \partial_\mu \bar{S}_1 \partial^\mu S_1 \\ 
&\quad - \left (\bar{\partial}_{z_2} \bar{S}_1 - \frac{i}{\sqrt 2} \bar{B}\bar{C} \delta^2(x_6,x_7) \right ) \left ( \partial_{z_2} S_1 + \frac{i}{\sqrt 2} BC \delta^2(x_6,x_7) \right ) \Bigg \}\\
+ \frac{1}{g_{\rm YM}^2} \int_{3\cap3'}\!\! \!\! d^2x \Bigg \{  - \bigg | \bigg (\partial&^\mu + i \frac{ A_V^\mu - A_{{V'}}^\mu}{2}\bigg ) B \bigg |^2 - \frac{|Q_3 - S_3|^2}{2} |B|^2 \\ 
&\quad - \left | \left (\partial^\mu - i \frac{ A_V^\mu - A_{{V'}}^\mu}{2}\right ) C\right |^2 - \frac{|Q_3 - S_3|^2}{2} |C|^2 \Bigg \} \,.
\end{split} \label{component}
\ee
Here $d^2x = dx^0\,dx^1$, while the indices $\mu, \nu$ run $0123$.  The 2-, 3-derivatives are set to zero by reduction, but terms with $A_V^{2,3}$ and $A_{V'}^{2,3}$ remain.  As usual, in the intersection action all brane fields are evaluated at the defect.\footnote{We are following the conventions of Ref.~\cite{Wess:1992cp}, where covariant derivatives contain a factor $\frac12$; the Abelian gauge fields, their scalar partners, and $g_{\rm YM}^2$ are normalized with respect to this. \label{wb2}}  The normalization of the intersection action can be changed by a scaling of $B$ and $C$; in the next section, where the action is noncanonical, this will no longer be true.
 
The action~(\ref{component}) is lengthy, but this is in part because its full symmetry is obscured.  First, we may check that $Q_1$ actually behaves like two components of the D3 gauge field, as desired.  Using $Q_1 = (iA_{V4} + A_{V5})/\sqrt{2} \equiv i A_{Vz_1}/\sqrt{2}$, the kinetic terms for $Q_1$ become
\be
- \frac{1}{4} \left ( \bar{\partial}_{z_1} Q_1 + \partial_{z_1} \bar{Q}_1 \right )^2 = - \frac{1}{2} \left ( \partial_{4} A_{V5} - \partial_{5} A_{V4} \right )^2 = - \frac{1}{2} F_{V45} F_V^{45}
\ee
and
\be
  -  \left | \partial^\mu Q_1 - \frac{i}{\sqrt{2}} \partial_1 A_V^\mu \right|^2 = - \frac{1}{2} \left ( F_{V\mu 4} F_V^{\mu 4} + F_{V\mu 5} F_V^{\mu 5} \right ) \,.
\ee

Additionally we may check that, after dimensional reduction, $A_{V2,3}$,  $A_{V'2,3}$, $Q_3$, and $S_3$ behave like transverse brane positions.  The fields $B$ and $C$ couple to these as
%\be
%- \frac{1}{4} \left ( |Q_3 - S_3|^2  + \left ( A_{V2} - A_{\mathcal{V}2} \right )^2 +  \left ( A_{V3} - A_{\mathcal{V}3} \right )^2 \right ) \left ( |B|^2 + |C|^2 \right )
%\ee
\be
- \frac{1}{2} \left ( |Q_3 - S_3|^2  +  |Q_0 - S_0|^2 \right ) \left ( |B|^2 + |C|^2 \right ) \,, \label{potent}
\ee
where we have defined
$Q_0 = (iA_{V2} + A_{V3} )/\sqrt 2$ and $S_0 =  (iA_{{V'}2} + A_{{V'}3} )/\sqrt 2$.   The separation between the branes produces a mass term for the $B$ and $C$ fields, from the tension of the stretched string.  The expected SO(4) symmetry is manifest in this form.  %{\color{red} I couldn't think of a better name for $Q_0$.  I changed the factor of 2 for esthetics, but haven't checked yet. }

 The action appears to contain quadratically divergent terms, such as
\be
- \frac{1}{8} \left (|C|^2 - |B|^2  \right ) \delta^2(0,0) \,, \quad  - \frac{1}{4} |B|^2 |C|^2 \delta^2(0,0) \,,
\ee
obtained after integrating over $z_1$ and $z_2$.  However, these automatically cancel.  They arise from the squares of the auxiliary fields
\begin{align}
D_V &= - \frac{\sqrt{2}}{2} \left ( \bar{\partial}_{z_1} Q_1 + \partial_{z_1} \bar{Q}_1 \right ) + \frac{1}{2} \left (|C|^2 - |B|^2  \right ) \delta^2(x_4,x_5) \,,\\
F_{Q_3} &= -  \bar{\partial}_{z_1} \bar{Q}_2 + \frac{i}{\sqrt2} \bar{B}\bar{C} \delta^2(x_4,x_5) \,,
\end{align}
and the corresponding terms on the other brane.
Energetics requires that the other term in each auxiliary field have a compensating delta function, and so the behavior of the fields near the brane must be
\be
Q_1 \approx \frac{1}{8 \pi \sqrt{2}} \frac{ |C|^2 - |B|^2 }{z_1} \,,\quad
Q_2 \approx - \frac{i}{4 \pi \sqrt2} \frac{BC}{{z}_1} \,. \label{nearbrane}
\ee
This removes the explicit squares of Dirac delta-functions, so the action is  free of quadratic divergences.  There are still logarithmic divergences, from the poles in the fields~(\ref{nearbrane}).  Their proper treatment via renormalization will be discussed in Section \ref{RenormalizationSection}.

The specific combination of intersection fields $BC$ sources the derivative of $Q_2$ in the $z_1$ direction and $S_1$ in the $z_2$ direction.  Noting that $Z_2 = \pi \alpha' Q_2/\sqrt 2$ is the collective coordinate for the D3-brane,\footnote{The usual relation between canonical brane scalars and collective coordinates is $X = 2\pi\alpha'\Phi$.  Here there is an extra $\frac12$ from the normalization convention noted in footnote~\ref{wb2}, and an extra $\frac{1}{\sqrt 2}$ from the normalization~(\ref{znorm}) of $z_2$.}
 we have
\be
Z_1 Z_2 = - \frac{i\alpha'}{4} BC \,.  \label{reconnect}
\ee
This is symmetric between the two branes, agreeing with the corresponding constraint from $F_{S_3}$.  Thus $BC$ is the modulus for the reconnection of the branes.

\subsection{Electric and magnetic sources}

The first three terms in the D3 part of the Lagrangian~(\ref{component}) can be nicely condensed into 
\be
-\frac{1}{4} {\cal F}_{ab} {\cal F}^{ab} \,.
\ee
Here $a,b$ run over the D3 directions 0145 and ${\cal F}_{ab}$ is identified with $F_{Vab}$ except for the component
\be
{\cal F}_{45} = D_V = \partial_{4} A_{V5} - \partial_{5} A_{V4}  + \frac{1}{2} \left (|C|^2 - |B|^2  \right ) \delta^2(x_4,x_5) \,.
\label{f45}
\ee
Similarly we define on the D3$'$ the tensor ${\cal F}'_{mn}$, with $m,n$ running over 0167 and with ${\cal F}'_{67} = D_{ V'}$.
The equations of motion for the gauge field are 
\be
\begin{split}
\partial_a {\cal F}^{ab} = & j^b_{E,V} = \frac{i}{2}  \left [ \left ( B \partial^b \bar{B} - \bar{B} \partial^b B \right ) - \left ( C \partial^b \bar{C} - \bar{C} \partial^b C \right ) \right ] \delta^2(x_4,x_5) \,,
\\
\partial_m {\cal F}^{\prime mn} = & j^n_{E,{V'}} = - \frac{i}{2}  \left [ \left ( B \partial^n \bar{B} - \bar{B} \partial^n B \right ) - \left ( C \partial^n \bar{C} - \bar{C} \partial^n C \right ) \right ] \delta^2(x_6,x_7) \,.
\end{split} \label{eom}
\ee
%{\color{red}(Check conventions on delta function conversion to real coordinates actually makes sense)} 
Note that only the 01 components of the currents are nonzero.

For the dual tensor one finds
\be
\begin{split}
\partial_a \tilde{\cal F}^{ab} = & j^b_{M,V} =   \frac{1}{2} \epsilon^{bc} \partial_c \left (|C|^2 - |B|^2  \right ) \delta^2(x_4,x_5) \,,
\\
\partial_m \tilde{\cal F}^{\prime mn} = & j^n_{M,{V'}} = - \frac{1}{2} \epsilon^{np} \partial_p \left (|C|^2 - |B|^2  \right ) \delta^2(x_6,x_7)\,. 
\end{split} \label{bianchi}
\ee
Here $\epsilon^{01} = - \epsilon^{10} = 1$ with other components vanishing, so the current is again tangent to the intersection.

It is the square of $\cal F$ that appears in the Lagrangian and the Hamiltonian, so it is natural to identify this with the field strength.  Eqs.~(\ref{eom}, \ref{bianchi}) thus represent electric and magnetic sources.  Indeed, this kind of action for an Abelian gauge field coupled to electric and magnetic charges is well-known~\cite{BNZ}, though in the present case there is the simplification that the magnetic couplings are local because the magnetic currents are topological (curls).
 In particular, the total magnetic charges are
\be
Q_M = -Q'_M = \frac{1}{2} \left( {\cal P}(x^1 = \infty) - {\cal P}(x^1 = -\infty) \right) \,, \label{qm}
\ee
where ${\cal  P} = |C|^2 - |B|^2$ is the contribution of the charged fields to the $D$-terms.  A 1+1 dimensional kink, with  values of ${\cal  P}$ differing at $\pm \infty$, will therefore carry a magnetic charge.

This somewhat surprising result, that the intersection hypermultiplets couple magnetically as well as electrically, is the first step toward identifying the D-string, but there is a complication.  At nonzero brane separation, the potential~(\ref{potent}) is proportional to $|C|^2 + |B|^2$ and takes its minimum value only when $B = C = {\cal P} = 0$.  A configuration of nonzero magnetic charge would necessarily have a nonzero potential energy density in at least one of the two asymptotic directions, and so an energy which diverges linearly at long distance.  The lagrangian~(\ref{component}) describes a system in which magnetic charges are confined.   This cannot account for the D-string, which exists as an isolated BPS particle.  In the next section we resolve this issue.

\section{Noncanonical  action}
\label{NonMinimalGaugeSection}
\setcounter{equation}{0}

We would obtain the expected spectrum, with unconfined monopoles of quantized charge, if the potential~(\ref{potent}) were of sine-Gordon form, periodic with a series of zeroes.  To see why this is the case, let us look again at the brane interpretation of the hypermultiplet fields.  We have noted that the product $BC$ sources brane bending through the $F_{Q_3}$ potential.  Similarly, ${\cal P}$ sources the vector potential as in Eq.~(\ref{nearbrane}),
\be
A_{Vz_1} \approx -i\frac{\cal P}{8\pi z_1} \,. \label{aclose}
\ee
This potential is locally pure gauge, but can have nontrivial holonomy.  In particular, when the brane intersection is resolved as in Eq.~(\ref{reconnect}), the point $z_1 = 0$ is pushed off to infinity and the remaining space is multiply connected.  The holonomy is
\be
\oint dz_1\, A_{Vz_1} + {\rm c.c.} = \frac12\,
{\rm Re}\,{\cal P} \,. \label{holo}
\ee
A shift of the holonomy by $4\pi$ should give a physically equivalent configuration, where the extra factor of 2 arises from the convention noted in footnote~\ref{wb2}.  Therefore, 
\be
{\cal P} \sim {\cal P} + 8\pi \label{pper}
\ee
must be a periodic variable.  With this, the magnetic charge~(\ref{qm}) has just the expected quantization.\footnote{Note that the allowed electric charges include not only the $(1,-1)$ of the intersection fields, but also $(\pm 1,0)$ and $(0,\pm 1)$ for strings that attach only to one brane and end elsewhere.}

We therefore consider more general actions, in which the hypermultiplets have arbitrary K\"ahler potentials.  This might have been anticipated, since the $B$ and $C$ fields on the 1+1 dimensional intersection are dimensionless.  At the end of this section, we will return to the $T$-dual higher dimensional configurations discussed in the introduction.
 
\subsection{General K\"ahler potentials}
Gauge invariance and manifest $\mathcal{N}=1$ supersymmetry restrict our general intersection action to be written in terms of a K\"{a}hler potential $K(B e^{{V}-V'}, \bar{B} , Ce^{{V'}-V}, \bar{C}   )$ and a superpotential of the form
\be
\begin{split}
W = \frac{i}{\sqrt 2} (Q_3 - S_3) f(BC)
\end{split}
\ee
for some function $f$.   Enforcing $\mathcal{N}=2$ supersymmetry additionally requires the K\"{a}hler metric to be Ricci flat and fixes $f$ in terms of $K$.  The form for $f$ is most easily determined from the component form of the action by requiring that the SO(4) symmetry associated with the $\mathcal{N}=2$ supersymmetry remains preserved.
 
In order to write the action concisely in component form, collect the fields $B$ and $C$ into a single vector $Z^\alpha = (B,C)$.   Define the Killing vector gauged by the brane gauge fields as $k^\alpha = (-iB,iC)$.  Additionally, we make use of the relation
\be
\partial_\alpha f(BC) = i \epsilon_{\alpha \beta} k^\beta f^\prime(BC)
\ee
%\be
%g^{\alpha \dot{\beta}} \frac{\partial W}{\partial Z^\alpha} \frac{\partial \bar{W}}{\partial \bar{Z}^{\dot{\beta}}}
%\ee
The bosonic sector of the component form of the generalized action may be written
\be
\begin{split}
S  = \frac{1}{g_{\rm YM}^2} \int_{\rm D3}\!\! d^2 x\, d x_4 dx_5 \Bigg \{&  - \frac{1}{4} F^{\mu \nu}_V F_{V \mu \nu} - \frac{1}{4} \left ( \bar{\partial}_{z_1} Q_1 + \partial_{z_1} \bar{Q}_1- \frac{1}{\sqrt{2}} {\cal P} \delta^2(x_4,x_5  )\right )^2  \\ 
&\quad  -  \left | \partial^\mu Q_1 - \frac{i}{\sqrt{2}} \partial_{z_1} A_V^\mu \right|^2    - \partial_\mu \bar{Q}_2 \partial^\mu Q_2  - \partial_\mu \bar{Q}_3 \partial^\mu Q_3  - \partial_1 Q_3 \bar{\partial}^1 \bar{Q}_3 \\ 
& \quad- \left (\bar{\partial}_{z_1} \bar{Q}_2 - \frac{i}{\sqrt 2} \bar{f} \delta^2(z_1, \bar{z}_1) \right ) \left ( \partial_{z_1} Q_2 + \frac{i}{\sqrt 2} f \delta^2(z_1, \bar{z}_1) \right ) \Bigg \} \\
+  \frac{1}{g_{\rm YM}^2} \int_{\rm D3'}\!\!\!  d^2 x\, d x_6 dx_7 \Bigg \{&  - \frac{1}{4} F^{\mu \nu}_{{V'}}F_{{V'} \mu \nu} - \frac{1}{4} \left ( \bar{\partial}_{z_2} S_2 + \partial_{z_2} \bar{S}_2 + \frac{1}{\sqrt{2}} {\cal P} \delta^2(z_2,\bar{z}_2)\right )^2  \\ 
&\quad  -  \left | \partial^\mu S_2 - \frac{i}{\sqrt{2}} \partial_{z_2} A_{{V'}}^\mu \right|^2   - \partial_{z_2} S_3 \bar{\partial}_{z_2} \bar{S}_3 - \partial_\mu \bar{S}_3 \partial^\mu S_3   - \partial_\mu \bar{S}_1 \partial^\mu S_1 \\ 
&\quad - \left (\bar{\partial}_{z_2} \bar{S}_1 - \frac{i}{\sqrt 2} \bar{f} \delta^2(z_2, \bar{z}_2) \right ) \left ( \partial_{z_2} S_1 + \frac{i}{\sqrt 2} f \delta^2(z_2, \bar{z}_2) \right ) \Bigg \}\\
+ \frac{1}{g_{\rm YM}^2} \int_{3\cap3'}\!\! \!\! d^2x \Bigg \{   - g_{\alpha \dot{\beta}}\,\,&\!\! \bigg ( \partial^\mu Z^\alpha + \frac{ A_V^\mu - A_{{V'}}^\mu}{2} k^\alpha \bigg ) \left ( \partial_\mu \bar{Z}^{\dot{\beta}} + \frac{ A_{V\mu} - A_{V'\mu}}{2} \bar{k}^{\dot{\beta}} \right ) \\ 
&\qquad\qquad\qquad - \frac{|Q_3 - S_3|^2}{2} g^{\alpha \dot{\gamma}}  \epsilon_{\alpha \beta} k^\beta \epsilon_{\dot{\gamma} \dot{\delta}} \bar{k}^{\dot{\delta}} |f^\prime|^2 \Bigg \} \,.
\end{split} \label{gencomponent}
\ee
Here $g_{\alpha \dot{\beta}} = \partial_\alpha \partial_{\dot\beta} K$ is the K\"{a}hler metric, and $\mathcal{P}$ is the moment map of the Killing vector $k^\alpha$, 
\be
\mathcal{P} = i k^\alpha \partial_\alpha K \,. \label{calp}
\ee
Note that Eqs.~(\ref{qm},$\,$\ref{aclose},$\,$\ref{holo}) continue to hold, but the relation between $\cal P$ and $B, C$ is modified in general.  Note also that the definition of $\mathcal{P}$ includes only the variations of $B , C$ and not the term from the variation of $Q_1$.

It is now straightforward to determine the function $f$.  The SO(4) symmetry mixes $Q_3$ into $A_{V2}$ and $A_{V3}$, so demanding these terms appear identically in the action sets
\be
k^\alpha g_{\alpha \dot{\beta}} \bar{k}^{\dot{\beta}} = g^{\alpha \dot{\gamma}}  \epsilon_{\alpha \beta} k^\beta \epsilon_{\dot{\gamma} \dot{\delta}} \bar{k}^{\dot{\delta}} |f^\prime|^2
\ee
Since the K\"{a}hler metric is just a two by two matrix, it can be inverted explicitly to give the relationship $g^{\alpha \dot{\gamma}} = (\det g)^{-1} \epsilon^{\alpha \beta} \epsilon^{\dot{\gamma} \dot{\delta}} g_{\beta \dot{\delta}} $, which yields
\be
|f^\prime|^2 = \det g \,.
\label{fConstraint}
\ee
The condition that the K\"{a}hler metric be Ricci flat is 
\be
0 = R_{\alpha \dot{\beta}} = - \partial_\alpha \partial_{\dot{\beta}} \ln \left ( \det g \right ) \,,
\ee
which is equivalently that the requirement that $\det g$ split into a holomorphic and anti-holomorphic part.  Therefore, for any K\"{a}hler potential preserving $\mathcal{N}=2$ supersymmetry, we are guaranteed to be able to solve Eq.~\ref{fConstraint} to find a superpotential that also preserves the supersymmetry.
 
As an aside, it is worth noting that to make the preservation of supersymmetry apparent in the fermionic sector, it is necessary to also make a field redefinition of the fermion fields.  For the vector multiplet fermion $\lambda_V$, the $Q_3$ chiral multiplet fermion $\psi_{Q_3}$, and the $Z^\alpha$ fermions $\psi_Z^\alpha$, we expect $(\lambda_V, \psi_{Q_3})$ and $(\psi_Z^1, \bar{\psi}_Z^2)$ to transform as $SU(2)$ doublets.  However, the relevant Yukawa couplings
\be
\sqrt{2} \bar{\lambda}_V \psi_Z^\alpha g_{\alpha \dot{\beta}} \bar{k}^{\dot{\beta}} \quad \quad \quad -i \sqrt{2} \bar{\psi}_{Q_3} \bar{\psi}_Z^\alpha \bar{\partial}_\alpha \bar{f}
\ee
are not singlets under this $SU(2)$ transformation.  By demanding the new fields obey this condition, the necessary field redefinition takes the form
\be
\psi_Z^{\prime \alpha} =  -i \left ( k^\beta g_{\beta \dot{\gamma}} \bar{k}^{\dot{\gamma}} \right )^{-1/2}\bar{f}^\prime  \epsilon^{\alpha \sigma} \left (g_{\sigma \dot{\lambda}} \bar{k}^{\dot{\lambda}} \psi^B_Z - \partial_\sigma f \psi^C_Z \right ) \,.
\ee
This gives the Yukawa couplings
\be
\begin{split}
& \bar{\lambda}_V \psi_Z^{\prime B}\sqrt{2k^\delta g_{\delta \dot{\lambda}} \bar{k}^{\dot{\lambda}}}
\\ & \bar{\psi}_{Q_3} \bar{\psi}^{\prime C}_{Z} \sqrt{2k^\delta g_{\delta \dot{\lambda}} \bar{k}^{\dot{\lambda}}}
\\ & \lambda_V \bar{\psi}^{\prime C}_{Z} \cdot 0
\\ & \psi_{Q_3} \psi^{\prime B}_{Z} \cdot 0
\end{split}
\ee
Thus in order to preserve $\mathcal{N}=2$ supersymmetry in the fermion sector, half the Yukawa couplings must vanish.
%{\color{red} I think it's good to put this in, someone may find it useful.  But no need to go further. }

We need a K\"ahler potential which satisfies the conditions for ${\cal N}=2$ supersymmetry and for which ${\cal P}$ is periodic.  In \S3.2 we will construct this explicitly, but first we can deduce how the periodic identification must act on the $(B,C)$ fields.  First, we claim that the function $f(BC)$ must remain equal to $BC$.  The product $BC$ transforms as $(+1,+1)$ under the $SO(2)_{45} \times SO(2)_{67}$ $R$-symmetries, so a more general superpotential would not transform properly.
Then the periodicity must be
\be
B \to B/(\beta BC)\,, \quad C \to C(\beta BC)  \label{bcperiod}
\ee
for some constant $\beta$.
First, this preserves the holomorphicity and gauge quantum numbers of the fields.  Second, it leaves invariant $BC = f$, which has a physical interpretation~(\ref{reconnect}) in terms of the brane configuration.  Finally, this identification would seem not to commute with $SO(2)_{45} \times SO(2)_{67}$.  However, the shift of the $A_{Vz_1}$ holonomy requires a gauge transformation $e^{i\theta} = \sqrt{z_1/\bar z_1}$.  This adds a gauge piece to the rotation generator in the $45$ plane, exactly cancelling the effect of the transformation~(\ref{bcperiod}); corresponding arguments apply in the $67$ plane.

\subsection{Gibbons-Hawking geometry}
\label{TaubNutSection}
We want to construct a specific K\"{a}hler metric that both preserves $\mathcal{N}=2$ supersymmetry and provides the correct periodicity for the topological charge $\mathcal{P}$.  The Gibbons-Hawking metrics provide a large class of $\mathcal{N}=2$ supersymmetric theories with in two complex dimensions, and we will see that there is a natural metric in of this type. %{\color{red}Will calling $\mathcal{P}$ the D-term confuse people? - A good point, I have added a note below Eq.~\ref{calp}}

The Gibbons-Hawking metrics are of the form
\be
ds^2 = V(\vec x) d\vec x \cdot d\vec x + \frac{1}{V(\vec x)}\omega^2  \,,
\ee
where $\vec x = (x,y,z)$ and
\begin{align}
\omega &= d\theta + \vec a(\vec x)\cdot d\vec x \,, \quad d\omega = da = *dV \,, \quad
\vec \partial \cdot \vec \partial V = 4\pi \sum_i \delta^3(\vec x - \vec x_i) \,.
\end{align}
To express these in K\"ahler form we follow Lebrun~\cite{lebrun91}, with the function~$u$ from that reference set to zero.  Define
\be
U = x + i y \,,\quad \phi = V dz + i \omega \,,
\ee
so that 
\be
ds^2 = V dU d\bar U + \frac{1}{V} \phi\bar \phi \,.
\ee
Then one finds that
\be
d\phi = d(x + iy) \wedge [(\partial_x - i \partial_y) V dz -  {\textstyle\frac12}\partial_z V d(x - iy)]  \,.
\ee
Since the right-hand side is closed it follows that locally we can write
\be
 V dz + i \omega = \phi = dW + \sigma dU \,, \label{defk}
\ee
where
\be
d\sigma = -(\partial_x  - i \partial_y) V dz + {\textstyle\frac12} \partial_z V d(x - iy) + O(d(x+iy)) \,. \label{ds}
\ee
We can regard Eqs.~(\ref{defk},$\,$\ref{ds}) as differential equations determining $W$ and $\sigma$.  Picking out the $dz$ components gives
\be
\partial_z W = V\,, \quad \partial_z\sigma = (-\partial_x  + i \partial_y )V\,. \label{diffe}
\ee
These hold in an axial gauge, $\omega_z = 0$.  The real part of Eq.~\ref{defk} is
\be
 2V dz = dW + d\bar W + \sigma dU + \bar\sigma d\bar U \,, \label{defk}
\ee
which determines the partial derivatives of $z$ with respect to the complex coordinates.\footnote{There are two natural sets of independent variables, the Gibbons-Hawking coordinates $(x,y,z,\theta)$ and the complex $(U, \bar U, W, \bar W)$.  Which we are using is determined by context: which derivatives appear in a given equation.  Also, when we differentiate with respect to $W_1$ we hold fixed $x,y,\theta$ or equivalently $U, \bar U, W_2$, so the change of variables between $z$ and $W_1$ is one-dimensional.}

Then $U$ and $W \equiv W_1 + i W_2$ are the appropriate complex coordinates on the space.  The metric is Hermitian,
\be
ds^2 = V dU d\bar U + \frac{1}{V} (dW + \sigma dU)(d\bar W + \bar\sigma d\bar U) \,. 
\ee
It is also K\"ahler, from $d\Omega = 0$ where
\be
\Omega = V dU\wedge d\bar U + \frac{1}{V} \phi\wedge \bar \phi = -2 i V dx \wedge dy  -2i dz \wedge \omega \,.
\ee
%Title page
We see that $\det g = 1$ so it is Ricci-flat. 

A special case is $V = 1/r$ where $r = |\vec x|$.  We have
\be
W = \sinh^{-1} \bigg(\frac{z}{\sqrt{x^2 + y^2}}\bigg) + i\theta \equiv W_1 + i W_2  \,,\quad   \sigma = \frac{z}{(x+iy) r}\,.
\ee
The metric in complex coordinates is then
\be
ds^2 = \sqrt{U\bar U} \cosh W_1 \left( dW d\bar W + \frac{ dU d\bar U}{U \bar U} \right)
+  \sqrt{U\bar U} \sinh W_1 \left( \frac{dW dU}{U} + \frac{ dW d\bar U}{\bar U} \right) \,.
\ee
This is obtained from the K\"ahler potential
\be
K = 4  \sqrt{U\bar U} \cosh W_1 \,.
\ee
This metric is actually flat: let 
\be
U = BC/2\,, \quad e^W = B/C\,,  \label{bcdef}
\ee
and then $K = B\bar B + C \bar C$.  In fact these identifications could have been anticipated, up to $c$-number normalization, in terms of the $U(1)$ symmetries by shifts of $\theta$ and rotations of $x + i y$.  Thus they continue to hold in the nonlinear case.

Returning to the general case, the $U(1)$ invariance on $\theta$, implies that the K\"aher potential $K$ is a function of $W_1$ but not $W_2$.  Then 
\be
\frac{1}{V} = g_{W\bar W} = \partial_W \partial_{\bar W} K = \frac14 \partial_{W_1}^2 K \,.
\ee
From~(\ref{diffe}), $\partial_z W_{1} = V $, and we can write
\be
4 = \partial_{W_1}^2 K \partial_z W_{1} = \partial_z \partial_{W_1} K \ \Rightarrow\ \partial_{W_1} K = 4(z + d(x,y)) \,.
\ee
Further,  
\be
 \frac{\bar\sigma}{V} = g_{\bar U W} = \partial_{\bar U} \partial_W K = \frac12 \partial_{\bar U} \partial_{W_1} K = 2  \partial_{\bar U} (z + d(x,y))
= \frac{\bar\sigma}{V} + 2  \partial_{\bar U} d(x,y) \,,
\ee
so that $d(x,y)$ is a constant that can be set to zero.  Noting that $k^\alpha \partial_\alpha W = k^B - k^C = -2i$,
the moment map is then
\be
{\cal P} = i k^\alpha \partial_\alpha \partial_W K = \partial_{W_1}K  = 4z  \,.  \label{cpz}
\ee

To obtain the periodicity~(\ref{pper}), Eq.~(\ref{cpz}) suggests that we should take
\be
V \stackrel{?}{=} \sum_{n=-\infty}^\infty \frac{1}{\sqrt{x^2 + y^2 + (z + 2\pi n)^2}} \,,
\ee
but this diverges.  We therefore regulate and subtract,
\be
V =  c + \lim_{N\to\infty} \left( -\frac{1}{\pi} \ln N + \sum_{n=-N}^N \frac{1}{\sqrt{x^2 + y^2 + (z + 2\pi n)^2}}  \right)\,.
\label{TaubNutSoln}
\ee
The parameter $c$ must be fixed by a stringy calculation, but its precise value will not matter.  For any $c$ the renormalized $V$ becomes negative for large enough $x^2 + y^2$, 
\be
V \sim c - \frac{1}{2\pi} \ln \frac{x^2 + y^2}{4\pi^2} \,.  \label{vsing}
\ee
This implies a singularity in the metric, whose implications we will discuss later.  The $U(1)$ symmetries are as before, and so the identification~(\ref{bcdef}) is the same.   

This metric first appeared in Ref.~\cite{Ooguri:1996me} as the metric on hypermultiplet moduli space near a conifold singularity, and in dual form in Ref.~\cite{Seiberg:1996ns,Gaiotto:2008cd} as the moduli space metric in a $U(1)$ gauge theory with one hypermultiplet on $R^3 \times S^1$.

The differential equation for $W$ implies that
\be
W = i\theta + cz + \lim_{N\to\infty} \left( -\frac{z}{\pi} \ln N + \sum_{n=-N}^N \sinh^{-1} \frac{z + 2\pi n}{\sqrt{x^2 + y^2}}  \right)
+ h(U,\bar U) \,.
\ee
for some function $h$.
Shifting by one period $z \to z+2\pi$ removes the $n=-N$ term and adds a term $n=N+1$, with the net effect
$W \to W + c\pi + \ln(16\pi^2/U\bar U)$.  This cannot be correct, as it does not respect the holomorphy, but we note that this can be repaired if the shift of $z$ is accompanied by a shift $\theta \to \theta - i \ln(\bar U/U)$.  The full periodicity is then
\be
W \to W + c\pi + \ln \frac{16\pi^2}{U^2} \, . \label{wperiod}
\ee
One could use as complex coordinates $U$ and $Y = W/\ln(U^2/16\pi^2 e^{\pi c})$, for which the periodicity is simply $(U,Y) \to (U, Y-2)$.
Using the relation~(\ref{bcdef}) of $W$ to $B,C$, the periodicity of $(B,C)$ is precisely~(\ref{bcperiod}), with $\beta = e^{-\pi c/2}/8\pi$.

\subsection{$T$-dual configurations}

We might have anticipated the need for a nonlinear kinetic term, since the fields $B,C$ are dimensionless in two dimensions.  However, we have noted that the soliton persists if we $T$-dual on $k$ DD directions, leading to a higher dimensional theory in which $B,C$ have units of $m^{k/2}$.  The point is that while the nonlinear terms are irrelevant in the renormalization sense, as two-derivative terms they are important on the moduli space.  Note that the D-string has mass $|Q_3|/g$ (for separation in the $z_3$-direction), but the D$(1+k)$ monopole in the $T$-dual theory has tension $|Q_3|(2\pi\alpha')^{-k/2}/g$, corresponding to a mass scale of order $|Q_3|^{1/(k+1)} \alpha'^{-k/2(k+1)}$.  This does go to zero with the separation, but it retains a dependence on the UV scale, which enters into the low energy theory through the couplings of the irrelevant terms.

For the D3-D5 theory shown in Table~2, the interpretation of the hypermultiplets is different.  The D5-brane bisects the D3 on the line $x^4 = x^5 = x^6 = x^7 = 0$.  The two halves of the D3 can be shifted independently in the 567 directions, since D3's can end on D5's.  The hypermultiplet moments correspond to the separation of the D3 endings along the D5~\cite{Hanany:1996ie}.  In this case the moduli space is $R^3$ and there is no periodicity.

\section{Renormalization}
\label{RenormalizationSection}
\setcounter{equation}{0}

We have noted that there will be logarithmic divergences in the classical action.  These arise in theories with codimension-two defects, because the fall-off of a point-sourced massless field in two dimensions is logarithmic.  Ref.~\cite{Goldberger:2001tn} addressed this issue, showing that these classical divergences fit naturally into the framework of renormalization theory.  In particular, they induce a flow of the effective action with scale.\footnote{This is also similar to the renormalization of the Schr\"odinger equation with a delta-function potential in 2+1 dimensions~\cite{Jackiw:1991je}.}

Procedurally, the renormalization is calculated no differently than in quantum field theory, except we work with tree level Feynman diagrams instead of loop diagrams.  Consider tree-level diagrams whose external lines are all $Z^\alpha$ or $\bar{Z}^{\dot{\alpha}}$.  Internal exchanges of $A_{V0,1}$, $Q_1$, or $Q_2$ can all lead to logarithmic divergences.  The tree-level vertices in diagrams for these fields are localized to the defect, and so their Fourier transforms are independent of the transverse momenta.  This means that the momentum of an internal line is unconstrained in two of the dimensions and must be integrated over, similar to the more usual loop integral over momentum.  This calculation may be done in either with superfields or with components and we will present both. %{\color{red} Is it good to present both?  I think it's good: each in its own section so people can skip over, but it's there if anyone needs it. Update: we need it, in sec. 5.}

\subsection{Superfield calculation}
Begin by introducing sources $J_{1,2,3}$ for the chiral fields $Q_{1,2,3}$ and $J_V$ for the vector field $V$.  First we integrate out $Q_{2,3}$.  We have
\be
S_{Q_{2,3}} = \int d^4x\, d^4\theta \left( \bar Q_i Q_i + Q_i J_i + \bar Q_i J_i\right)  + \int d^4x\, d^2\theta\,  Q_3 \partial_1 Q_2 + {\rm c.c. }\,.
\ee
Now shift, $Q_i = Q'_i + q_i$.  The terms linear in $Q'$ are
\begin{align}
S_{O(Q')} &= \int d^4x\, d^4\theta \,Q'_i (q_i  + J_i )  - \int d^4x\, d^2\theta\,  \epsilon_{ij} Q'_i \partial_1 q'_j  
\nonumber\\
&= \int d^4x\, d^2\theta \,  Q'_i \left[ \bar D^2 (q_i  + J_i )/4  -   \epsilon_{ij} \partial_1 q'_j \right] .
\end{align}
Here $\epsilon_{23} = 1$.  In the second line we have combined terms, noting that $\int d\theta = \partial_\theta$ implies that $\bar D^2 X/4 = \int d^2\bar\theta\, X$ for any $X$, up to total $x$ derivatives (there is an implicit $\frac14$ in the integral so that $\int d^2\theta \,\theta^2 = 1$).  The condition that the linear term vanish gives 
\be
\left[ \begin{array}{c} q_2 \\ \bar q_3  \end{array} \right] = \frac{1}{p^2} 
\left[ \begin{array}{cc}\bar D^4/4 & \partial_1 \\ -\partial_1 &  D^2/4  \end{array} \right]  \left[ \begin{array}{c} D^2 J_2/4 \\ \bar D^2 J_3/4 \end{array} \right]  \,.
\ee
Here we have gone to momentum space, and $p^2 = p_\parallel^2 + p_\perp^2 = p_\mu p^\mu + p_1 \bar p_1 $.  We only need the propagator for $Q_3$ so can set $J_2 = 0$.  The source-dependence of the action is then given as usual by $\frac12$ of the source terms with classical background inserted,
\be
S_{V,Q} \to \int \frac{d^4p}{(2\pi)^4} d^4\theta\, q_3 J_3 = \int \frac{d^4p}{(2\pi)^4} d^4\theta\, \bar J_3 \frac{D^2 \bar D^2}{16 p^2} J_3 \,.
\ee
The propagator is then
\be
\langle Q_3(p,\theta) \bar Q_3(p',\theta') \rangle = -i (2\pi)^4 \delta^4(p+p')  \frac{\bar D^2 D^2}{16 p^2} \delta^4(\theta - \theta') \,.
\ee

The interaction is
\begin{align}
S_{\rm int} = \frac{i}{\sqrt 2} \int d^2x \, d^2\theta \, Q_3 f +  {\rm c.c. }& = -\frac{i}{\sqrt 2}\int d^2x \, d^2\theta \, Q_3 \frac{\bar D^2 D^2}{16 p_\parallel^2} f +  {\rm c.c. } 
\\
&= -\frac{i}{\sqrt 2}\int d^2x \, d^4\theta \, Q_3 \frac{D^2}{4 p_\parallel^2} f +  {\rm c.c. } \,.
\end{align}
The graph from exchanging a $Q_3$ is then
\be
i \int \frac{d^4 p}{(2\pi)^4} f(p_\parallel) \frac{D^2 \bar D^2 D^2 \bar D^2}{512 p_\parallel^2 p_\parallel^2 p^2 } \bar f(-p_\parallel)
= \frac{i}{2} \int \frac{d^4 p}{(2\pi)^4} \frac{1}{p^2 }  f(p_\parallel)\bar f(-p_\parallel) \,.
\ee
We see that the integral over $p_\perp^2$ generates a log divergence, which is canceled by an $\bar f f$ correction to the K\"ahler potential.  The effective K\"ahler potential is given by integrating out $p_\perp$ between a UV cutoff $\Lambda$ and the scale $\mu$,\footnote{To extract the log divergence we do not need the details of the cutoff, but in fact a simple transverse momentum cutoff preserves both supersymmetry and gauge invariance.}
\be
i \Delta K = \frac{i}{8\pi} \ln \left ( \frac{ \Lambda^2 }{\mu^2} \right ) \bar f f \,.
\ee
This contributes $- \frac{1}{4\pi}\bar f f$ to $\mu \partial_\mu K$; the second brane makes an equal contribution.

Now we integrate out $V$ and $Q_1$.  The action is
\begin{align}
S_{V,Q_{1}} = \int d^4x\, d^2\theta\,&\frac14  W_V^\alpha W_{V\alpha} + {\rm c.c.} 
\nonumber\\
&+  \int d^4x\, d^4\theta\,  \left( \bar\partial_1 V \partial_1 V
- \sqrt 2(Q_1 \bar\partial_1 V + \bar Q_1 \partial_1 V) + \bar Q_1 Q_1 + J_VV) \right) \,.
\end{align}
With the gauge transformation $Q_1 \to Q_1 + \partial_1\Lambda$ we can go to the unitary gauge $Q_1 =0$, in which
\be
S_{V,Q_{1}} =   -\int d^4x\, d^4\theta\,  V (P_T \partial_\mu \partial^\mu +
 \bar\partial_1 \partial_1) V
\,.
\ee
We use the projection operators defined in Ref.~\cite{Wess:1992cp}, chap.~IX.  The propagator is then
\be
\langle V(p,\theta) \bar V(p',\theta') \rangle = -i (2\pi)^4 \delta^4(p+p') \left( \frac{P_T}{p^2} + \frac{P_1 + P_2}{p_\perp^2} \right)\delta^4(\theta - \theta') \,.
\ee

The first order coupling of the field $V$ is $V \delta_\Lambda K = - V\delta_{\bar\Lambda} K\equiv -V{\cal P}$, where $K$ is the hypermultiplet K\"ahler potential and $\delta_\Lambda$ is the holomorphic part of the gauge transformation.
At large $p_\perp$, where the divergence enters, the projection operators just add to unity.  We thus obtain the divergence
\be
-\frac{i}{4} \int \frac{d^4 p}{(2\pi)^4} \frac{1}{p^2 }  {\cal P}(p_\parallel)\bar {\cal P}(-p_\parallel) \,,
\ee
and so
\be
i \Delta' K= -\frac{i}{16\pi} \ln \left ( \frac{ \Lambda^2 }{\mu^2} \right ) {\cal P}^2 \,.
\ee
In all, 
\be
\mu \partial_\mu K = \frac{1}{4\pi}({\cal P}^2 - 2\bar f f) \,,
\ee
including the contribution of the second brane.

It can be shown that this renormalization preserves $d=4$, ${\cal N}=2$ supersymmetry.  We illustrate this with the simple case 
\be
K = \bar B B + \bar C C\,,\quad f = BC
\ee
at the initial point $\mu = \mu_0$.  Then ${\cal P}= (\bar CC - \bar BB)$ and 
\be
\mu \partial_\mu K = \frac{1}{4\pi}(\bar B^2 B^2 - 4 \bar B B \bar C C + \bar C^2 C^2)\,;
\ee
we are working to linear order in the flow, so quantities on the right are evaluated at $\mu_0$. 
The flow of the metric is 
\be
\mu \partial_\mu g_{B\bar B} = \frac{1}{\pi}(\bar B B - \bar C C) = - \mu \partial_\mu g_{C\bar C},
\ee
which is traceless.  So to this linear order the ${\cal N}=2$ condition $\det g = 1$ is preserved.

\subsection{Component calculation}
Consider first exchange of a $Q_2$.  The relevant interaction is
\be
-  \left (\bar{\partial}^1 \bar{Q}_2 - \frac{i}{\sqrt 2} \bar{f} \delta^2 (x_4,x_5) \right ) \left (\partial^1 Q_2 + \frac{i}{\sqrt 2}  f \delta^2 (x_4,x_5  ) \right ) \,.
\ee
This leads to a Feynman diagram equation of the form
\be
\frac{(-i)^3}{2} f^{[n]} \bar{f}^{[m]} \delta^2\left (\sum_{i=1}^{n+m} \vec{k}_i \right ) \int \frac{d^2p_{\perp}}{(2\pi)^2} \frac{p_{\perp}^2}{p_{\perp}^2 + p_{\parallel}^2} 
\ee
for external momenta $\vec{k}_i$, transverse (to the intersection space) internal momenta $\vec{p}_{\perp}$, and parallel internal momenta $\vec{p}_{\parallel}$. The expression $f^{[n]}$ corresponds to the coefficient of some term of $f$ with $n$ total fields.  The parallel momentum is set by total momentum on either side of the graph:
\be
\vec{p}_{\parallel} = \sum_{i=1}^n \vec{k}_{fi} = - \sum_{i=1}^m \vec{k}_{\bar{f}i} \,.
\ee
If we regulate the momentum integral by a cut-off $\Lambda$ and integrate, we get
\be
\frac{(-i)^3}{8\pi} f^{[n]} \bar{f}^{[m]} \delta^2\left (\sum^{n+m}_{i=1} \vec{k}_i \right ) \left ( \Lambda^2 - p^2_\parallel \ln \left ( \frac{ \Lambda^2 }{p^2_\parallel} \right )\right ) \,.
\ee
The quadratic divergence is cancelled by the explicit $|f|^2 (\delta^2(x_4,x_5  ))^2$ term in the action.  For the logarithmic term, a finite renormalization scale $\mu$ can be introduced, and then we can ignore the non-infinite piece, giving
\be
- \frac{(-i)^3}{8\pi} f^{[n]} \bar{f}^{[m]} \delta^2\left (\sum_{i=1} \vec{k}_i \right ) p^2_\parallel \ln \left ( \frac{ \Lambda^2 }{\mu^2} \right )\,.
\ee
Since we may write
\be
p^2_{\parallel} = - \left (\sum_{i=1}^n \vec{k}_{fi} \right ) \cdot  \left ( \sum_{i=1}^m \vec{k}_{\bar{f}i} \right )  \,,
\ee
the diagram is proportional to all pairs of momenta, one taken from $f$ and one from $\bar{f}$.  This is equivalent to the contribution that would come from a Lagrangian term of the form
\be
-\frac{1}{8\pi} \ln \left ( \frac{\Lambda^2}{\mu^2} \right ) \partial^\mu Z^\alpha \left ( \partial_\alpha f \right ) \left ( \partial_{\dot{\beta}} \bar{f} \right ) \partial_\mu \bar{Z}^{\dot{\beta}} \,.
\ee
Since the K\"{a}hler potential enters the Lagrangian as $\partial^\mu Z^\alpha \left ( \partial_\alpha \partial_{\dot{\beta}} K \right ) \partial_\mu \bar{Z}^{\dot{\beta}}$, we can conclude that the correction to the K\"{a}hler potential from an internal $Q_2$ is
\be
\Delta K = \frac{|f|^2}{8 \pi} \ln \left (\frac{\Lambda^2}{\mu^2} \right ) \,.
\ee
Now we calculate the remaining contributions to the renormalization.  Since $Q_1$ is really the gauge field, consider the contribution from it and $A_{V0,1}$ together.  First rewrite the $Q_1$ interaction
\be
- \frac{1}{4} \left ( \bar{\partial}^1 Q_1 + \partial^1\bar{Q}_1 - \frac{1}{\sqrt{2}} \mathcal{P} \delta^2(x_4,x_5  ) \right )^2
\ee
in terms of $A_{4}$ and $A_5$ using $Q_1 = (iA_4 + A_5)/\sqrt{2}$ and $\partial_1 = \partial_4 - i \partial_5$ to get
\be
- \frac{1}{4} \left ( \frac{2}{\sqrt{2}} \partial_4 A_5 - \frac{2}{\sqrt{2}} \partial_5 A_4 - \frac{1}{\sqrt{2}} \mathcal{P} \delta^2(x_4,x_5  ) \right )^2 \,.
\ee
This leads to the usual $F_{45}^2/2$ kinetic term for the gauge field, as well as giving the interactions
\be
- \frac{1}{2} \left ( A_5 \mathcal{P} \partial_4 \delta^2(x_4,x_5  ) - A_4 \mathcal{P} \partial_5 \delta^2(x_4,x_5  )\right ) \,.
\ee
The $A_{0,1}$ couple to $(1/2)k^\alpha g_{\alpha \dot{\beta}} \partial_\mu \bar{Z}^{\dot{\beta}}=(-i/2)\partial_{\dot{\beta}} \mathcal{P} \partial_\mu \bar{Z}^{\dot{\beta}}$ and its complex conjugate, so we can now write a valid but unusual source for $A_\mu$ as
\be
J^\mu = \left ( \begin{array}{c} -\frac{i}{2} \partial_{\dot{\beta}} \mathcal{P} \partial^0 \bar{Z}^{\dot{\beta}} \delta^2(x_4,x_5  ) + {\rm c.c.} \\
 -\frac{i}{2} \partial_{\dot{\beta}} \mathcal{P} \partial^1 \bar{Z}^{\dot{\beta}} \delta^2(x_4,x_5  ) + {\rm c.c.} \\
  - \frac{1}{2} \mathcal{P} \partial^5 \delta^2(x_4,x_5  ) \\ \frac{1}{2} \mathcal{P} \partial^4 \delta^2(x_4,x_5  ) \end{array} \right) 
\ee
for $\mu=0,1,4,5$.  We can then use the usual $-i\eta_{\mu \nu}/(p_{\perp}^2 +p_{\parallel}^2)$ propagator for $A_\mu$ and calculate $J^\mu \Delta_{\mu \nu} J^\nu$.  At this point, it is easier to consider the $0,1$ and $4,5$ terms separately.  Starting with the $4,5$ terms together, we get a $p_{\perp}^2$ in momentum space from the $\partial^2_4$ and $\partial^2_5$.  The calculation is then almost identical to the previous case for $Q_2$.  Regulate and then integrate to get
\be
\frac{(-i)^3}{32\pi} \mathcal{P}^{[n]} \mathcal{P}^{[m]} \delta^2\left (\sum^{n+m}_{i=1} \vec{k}_i \right ) \left ( \Lambda^2 - p^2_\parallel \ln \left ( \frac{ \Lambda^2 }{p^2_\parallel} \right )\right ) \,.
\ee
%{\color{red} The quadratic piece then completely fails to be cancelled by the Dirac delta squared term in the Lagrangian.  But ignoring that for now and moving on:} Once the renormalization scale is introduced and the finite piece is dropped, we get
The quadratic piece is cancelled by $\delta^2(0)$ as expected.  Introducing the RG scale in place of $p$, the corresponding term in the effective Lagrangian is
%\be
%- \frac{(-i)^3}{16\pi} \mathcal{P}^{[n]} \mathcal{P}^{[m]} \delta^2\left (\sum^{n+m}_{i=1} \vec{k}_i \right ) p^2_\parallel \ln \left %( \frac{ \Lambda^2 }{\mu^2} \right ) \,.
%\ee
%This would be produced by a term in the Lagrangian of the form
\be
-\frac{1}{32\pi} \ln \left ( \frac{\Lambda^2}{\mu^2} \right ) \left (\partial_\mu Z^\alpha  \partial^\mu Z^\beta  \partial_\alpha \mathcal{P}  \partial_\beta \mathcal{P}   +2\partial_\mu Z^\alpha  \partial^\mu \bar{Z}^{\dot{\beta}} \partial_\alpha \mathcal{P}   \partial_{\dot{\beta}} \mathcal{P}    + \partial_\mu \bar{Z}^{\dot{\alpha}}  \partial^\mu \bar{Z}^{\dot{\beta}} \partial_{\dot{\alpha}} \mathcal{P}   \partial_{\dot{\beta}} \mathcal{P}   \right ) \,,
\ee
where the extra terms have arisen since $\mathcal{P}$ is not holomorphic, unlike $f$.  This term does not appear to preserve the K\"{a}hler structure of the metric, but we should first combine it with contribution from $A_{V0,1}$ before handling this issue.  
 
The contribution from an internal $A_{V0,1}$ is more straightforward.  The internal integral yields only a logarithmic divergence as $A_{V0,1}$ already couples to the form $\mathcal{P} \partial^\mu \bar{Z}^{\dot{\beta}}$.  We will then get
\be
-\frac{1}{32\pi} \ln \left ( \frac{\Lambda^2}{\mu^2} \right )  \left (\partial_\mu Z^\alpha  \partial^\mu Z^\beta  \partial_\alpha \mathcal{P}  \partial_\beta \mathcal{P}   - 2\partial_\mu Z^\alpha  \partial^\mu \bar{Z}^{\dot{\beta}} \partial_\alpha \mathcal{P}   \partial_{\dot{\beta}} \mathcal{P}    + \partial_\mu \bar{Z}^{\dot{\alpha}}  \partial^\mu \bar{Z}^{\dot{\beta}} \partial_{\dot{\alpha}} \mathcal{P}   \partial_{\dot{\beta}} \mathcal{P}   \right ) \,,
\ee
where the minus sign arises from the leading $i$'s in the vertex.  These combine to give
\be
-\frac{1}{16\pi} \ln \left ( \frac{\Lambda^2}{\mu^2} \right ) \left (\partial_\mu Z^\alpha  \partial^\mu Z^\beta  \partial_\alpha \mathcal{P}  \partial_\beta \mathcal{P}   
 + \partial_\mu \bar{Z}^{\dot{\alpha}}  \partial^\mu \bar{Z}^{\dot{\beta}} \partial_{\dot{\alpha}} \mathcal{P}   \partial_{\dot{\beta}} \mathcal{P}   \right ) \,.
\label{nonHermitianForm}
\ee
This is still not in the form of a hermitian metric.  In order to restore the K\"{a}hler form, we need to renormalize the fields as well.  Make the coordinate transform $Z^\alpha \to Z^\alpha + \xi^\alpha$ with
with
\be
\xi^\alpha = - \frac{1}{16\pi} \ln \left ( \frac{\Lambda^2}{\mu^2} \right ) \mathcal{P} \partial^\alpha \mathcal{P} \,.
\ee
The metric transforms as
\be
\begin{split}
\delta g_{\alpha \beta} &= \nabla_\alpha \xi_\beta + \nabla_\beta \xi_{\alpha}
\\ \delta g_{\alpha \dot{\beta}} &= \nabla_\alpha \bar{\xi}_{\dot{\beta}} + {\nabla}_{\dot{\beta}} \xi_{\alpha}\,,
\end{split}
\ee
which corrects Equation \ref{nonHermitianForm} to give
\be
\frac{1}{16\pi} \ln \left ( \frac{\Lambda^2}{\mu^2} \right ) \partial_\mu Z^{\alpha} \partial^\mu \bar{Z}^{\dot{\beta}}   \partial_\alpha \partial_{\dot{\beta}} \mathcal{P}^2 
\ee
or
\be
\Delta K = - \frac{1}{16\pi} \mathcal{P}^2 \ln \left ( \frac{\Lambda^2}{\mu^2} \right ) \,.
\ee
Combining the with the results from an internal $Q_2$ and doubling to include contributions from both branes gives the overall beta functions
\begin{align}
& \mu \partial_\mu K = \frac{1}{4\pi} \left ( \mathcal{P}^2 -2 |f|^2 \right ) \,,
\\ & \mu \partial_\mu Z^\alpha = - \frac{i}{4\pi} k^\alpha \mathcal{P} \,.  \label{fieldren}
\end{align}
Here we have used $\partial^\alpha {\cal P} = g^{\alpha\dot\beta} \partial_{\dot\beta}( i k^\gamma \partial_\gamma K) = ik^\alpha$.

The nonholomorphic renormalization of $Z^\alpha$ arises because of the nonlinearity of the supersymmetry transformation in Wess-Zumino gauge.  In the superfield gauge used in  \S4.1 holomorphy is preserved.

\subsection{Renormalized Gibbons-Hawking potential}

With the RG flow of the K\"{a}hler potential in hand, we can ask how it affects the specific K\"{a}hler potential given by Eq.~(\ref{TaubNutSoln}).  In terms of the Gibbons-Hawking coordinates the flow is
\be
\mu \frac{ \partial K}{\partial \mu}\bigg|_{W,\bar W, U, \bar U} = \frac{2}{\pi} (2z^2 - x^2 - y^2) \,.
\ee
%{\color{red}I think this is right, but there are some 2's to debug.}
It is important to be careful that the derivative is taken with the chiral fields held fixed.  Then using $ g_{W\bar W} = \partial_W \partial_{\bar W} K = 1/V$ and the earlier result $\partial_{W_1} z = 1/V$,
\be
\mu \frac{ \partial (1/V)}{\partial \mu}\bigg|_{W,\bar W, U, \bar U} = \frac{1}{\pi V^2} \partial_{W_1}^2( z^2) = \frac{2}{\pi V^3} (V - z \partial_{z}V)
 \,.
\ee
Now
\be
\mu \frac{ \partial }{\partial \mu}\bigg|_{W,\bar W, U, \bar U} = \mu \frac{ \partial }{\partial \mu}\bigg|_{x,y,z,\theta}
+ \mu \frac{ \partial z}{\partial \mu}\bigg|_{W,\bar W, U, \bar U} \frac{\partial}{\partial z} \,.
\ee
Using the earlier $z = \partial_W K/2$ gives
\be
\mu \frac{ \partial z}{\partial \mu}\bigg|_{W,\bar W, U, \bar U} = \frac{2}{\pi} \partial_W (z^2) = \frac{2z}{\pi V} \,,
\ee
and so
\be
\mu \frac{ \partial V }{\partial \mu}\bigg|_{x,y,z,\theta} = \mu \frac{ \partial V}{\partial \mu}\bigg|_{W,\bar W, U, \bar U}
- \frac{2z \partial_z V}{\pi V} = -\frac{2}{\pi} \,.
\ee

Thus the whole effect of the flow for any metric is on the constant term in the Gibbons-Hawking potential, in such a direction that it becomes more positive in the \mbox{IR}.  This partly offsets the singularity~(\ref{vsing}) found previously: in combination the two give
\be
V(\mu) \sim -\frac{1}{\pi} \ln\left[ \alpha' \mu^2 \sqrt{x^2 + y^2} \right] \,. \label{asymv}
\ee
One consequence is that the metric on moduli space, give by the $\mu \to 0$ limit, is positive and in fact infinite.  In other words, the moduli are frozen due to IR divergences, a phenomenon encountered previously in Ref.~\cite{Witten:1997sc}.  Thus the renormalization removes one of the potentially unphysical properties of our field theory.  

The metric on field space still breaks down if $x^2 + y^2$ is too large, for given $\mu$.  This suggests a simple interpretation: if we probe the resolved intersection on scales shorter than the resolution, then the effective field theory of the resolution in terms of the fields $B$ and $C$ breaks down and we must use the full DBI brane action.   Similarly, at fixed $x^2 + y^2$ there is a Landau pole in the UV. Thus the complete formulation of the intersection requires embedding in string theory.  In Refs.~\cite{Seiberg:1996ns,Gaiotto:2008cd} the same Landau pole appears, arising from a 3+1 gauge theory. 

Refs.~\cite{Cottrell:2014ura} have noted an alternate UV completion, the $SU(2)$ ${\cal N}=4$ gauge theory with a scalar v.e.v.\ that is linear in space and vanishes on a line.  One can think of this as the field theory on D3-branes intersecting at an angle, where the angle and $\alpha'$ are taken to zero toegether.

If we work at finite $Q_3$, we would expect the fluctuations of the $B,C$ fields to be of order $-g_{\rm YM} \ln \alpha'|Q_3|$, and the product of this with $\mu \sim |Q_3|$ can be made parametrically small.  Thus there should be that there is a regime in which the effective field theory can be quantized, and $S$-duality studied in the field theory.

The UV incompleteness is not special to the Gibbons-Hawking metric but occurs for the canonical metric as well.  In Gibbons-Hawking parameterization this is simply $V = 1/|\vec x|$ to leading order, and becomes
\be
V = \frac{1}{|\vec x|} - \frac{1}{\pi} \ln (\mu^2/\mu_0^2) \,.
\ee
Again there is a Landau pole.

\subsection{DBI action}

When $z_1 z_2 \gg \alpha'$, the brane curvature is small and the DBI action should be an effective description of the system.  This gives us an alternate means of calculating the moduli space metric, which should agree with that found above.
Thus we insert into the DBI action the $t,x$-dependent
\begin{equation}
z_1 z_2 = -i \frac{\alpha^\prime}{4} f(t,x) \, ,
\end{equation}
where in the field theory $f(t,x) =BC$ is a function of the intersection fields.  With all other fields zero, we will assume this form in the DBI action and find the resulting effective action for $f(t,x)$.
\paragraph*{}
With only metric fields turned on, the D3-brane DBI action is just
\begin{equation}
- \frac{T_p}{g_c}\int d^{4} \zeta \sqrt{|\det G_{ab}|} \, ,
\end{equation}
where $G_{ab}$ is the induced metric on the brane given in the coordinates $\zeta^a$.  A convenient choice of coordinates is $\zeta^a = (t,x,w_1,w_2)$ with
\begin{equation}
z_1 = \sqrt{\alpha^\prime} e^{w_1 + iw_2} \quad \quad z_2 =  i \frac{\sqrt{\alpha^\prime}}{4} f(t,x) e^{-w_1 - i w_2} \, .
\end{equation}
%In these coordinates, the induced metric on the brane is 
%\begin{equation}
%\begin{split}
%&G_{w\bar{w}}  = \alpha^\prime e^{i(w-\bar{w})} + \frac{\alpha^{\prime}}{8} e^{-i(w-\bar{w})} |f|^2
%\\ &G_{w \mu}  = - \frac{i \alpha^{\prime}}{8}e^{-i(w-\bar{w})} f \partial_\mu \bar{f}
%\\ &G_{\mu \nu}  = \eta_{\mu \nu} +\frac{\alpha^{\prime}}{8} e^{-i(w-\bar{w})} \partial_{(\mu} f \partial_{\nu)} \bar{f}
%\end{split}
%\end{equation}
%with $\mu=t,x$.  As expected, there is a symmetry in the metric in the angular direction $\mathrm{Re} (w)$, which is just the phase of $z_1$.  Writing $w_I = \mathrm{Im}(w)$, the action is then
The DBI action is then
\begin{equation}
S = - \frac{8\pi T_p \alpha^{\prime}}{g_c} \int dt\, dx\, dw_1 \left |\left [  \left (  e^{2w_1} +  \frac{1}{16} e^{-2w_1} |f|^2 \right ) +\frac{\alpha^\prime}{8} \partial_\mu f \partial^\mu \bar{f} \right ]^2 - \frac{\alpha^{\prime 2}}{64} \left |  \partial_\mu f \partial^\mu f \right |^2 \right |^{1/2} \,,
\end{equation}
with $\mu = t,x$.  As expected, dependence on the phase of $z_1$ drops out of the metric, and the integral in the $w_2$ direction has  been done.  The remaining integral in $w_1$ is divergent as $w_1 \to \pm \infty$, corresponding to contributions far out on the branes in the $z_1$ and $z_2$ directions.  This corresponds to an IR divergence associated with the infinite surface area of the extended D-branes, and should ultimately be $f$-independent.  To regulate symmetrically, set a cut-off at $|z_1| = |z_2| = L$, which restricts the integral to
\begin{equation}
-   \frac{1}{2} \ln \left ( \frac{16 L^2}{\alpha^\prime |f|^2} \right )   < w_1 < \frac{1}{2} \ln \left ( \frac{L^2}{\alpha^\prime} \right ) \, .
\end{equation}
%Even regulated, the integral in $w_I$ has no straightforward closed-form solution.  However, the $w_I$ term in the square root has a minimum value given by
%\begin{equation}
%\frac{|f|}{\sqrt{2}} \leq \left (  e^{2w_1} +  \frac{1}{8} e^{-2w_1} |f|^2 \right ) \, .
%\end{equation}

To find the effective action we expand in derivatives.
% may make the assumption that $f$ varies slowly with $(t,x)$ such that $\alpha^\prime | \partial_i f|^2 \ll |f|$ and then expand the square root.  
With only 8 supersymmetries, we expect the DBI action to match the gauge theories for terms with two or fewer derivatives, so expand to second order in derivatives of $f$, integrate out $w_1$, and drop terms that vanish for $L^2/\alpha^\prime \to \infty$ to get the effective action for $f$:
\begin{equation}
S = - \frac{8\pi T_p}{g_c} \int dt dx  \left \{ L^2+ \frac{\alpha'^2}{8} \ln \left ( \frac{4L^2}{\alpha^\prime |f|} \right ) \partial_\mu f \partial^\mu \bar{f}  \right \} \, . \label{dbiresult}
\end{equation}
The potential term is independent of $f$, consistent with the masslessness of the deformation modes.  The cutoff $L$ should be identified with $\mu^{-1}$.  For example, if we consider ripples of wavelength $\lambda$ then $L \sim \lambda$.
Noting that $f = BC = 2 (x + iy)$, we see that metric for $f$ corresponds to a Gibbons-Hawking potential 
\be
V(\mu) \propto {\rm constant} - \ln\left[ \alpha' \mu^2 \sqrt{x^2 + y^2} \right] \,.
\ee
This is precisely as found in Eq.~(\ref{asymv}) from the combination of classical renormalization and Gibbons-Hawking gymnastics, confirming our picture.

The DBI action gives a result that corresponds to the smearing of the Gibbons-Hawking sources into a line source, independent of the GH coordinate $z$.  This action cannot be sensitive to $z$ because this corresponds to a pure gauge excitation.  The leading effect that would be sensitive to $z$ would be string disk instantons: the resolution of the singularity pushes the point $z_1 = 0$ to infinity, producing a nontrivial circle which can bound a disk.  The world-sheet action Re$\,I \propto |z_1z_2|/2\pi\alpha' \propto |BC|$, Im$\,I = z$, is of the correct parametric form.

The full DBI action is stable, but the action~(\ref{dbiresult}) has the wrong sign for perturbations of sufficiently short wavelength.  For these the higher-derivative terms must be included in order to have a stable configuration.  This will result in the breakdown of the associated effective field theory, as argued in \S4.3.

%The effective field theory thus accurately predicts its own demise, and we have a consistency check on the renormalization as calculated in field theory.
%\begin{equation}
%S = - \frac{2\pi T_p \alpha^\prime }{g_c} \int dt dx (2 dw_I) \left ( e^{-2w_I} + \frac{1}{8} e^{2w_I} |f|^2 \right ) - \frac{\alpha^\prime}{8} \left ( |\partial_{t} f|^2 - |\partial_{x} f|^2\right) - \frac{\alpha^{\prime2}}{128} \frac{\left | \left( \partial_x f \right )^2 - \left( \partial_t f \right )^2 \right |^2 }{\left ( e^{-2w_I} + e^{2w_I} |f|^2/8 \right )}
%\end{equation}
%This integrates to
%\begin{equation}
%\begin{split}
%S = - \frac{2\pi T_p \alpha^\prime}{g_c} \int dt dx & \Bigg \{ \left (2\frac{L^2}{\alpha^\prime} - \frac{\alpha^\prime |f|^2}{4 L^2} \right ) - \frac{\alpha^\prime}{8} \ln \left ( \frac{2\sqrt{2}L^2}{\alpha^\prime |f|} \right ) \left ( |\partial_{t} f|^2 - |\partial_{x} f|^2\right) \\ & - \frac{\alpha^{\prime2}}{32\sqrt{2}} \left (\tan^{-1} \left (\frac{ L^2}{|f|} \right ) - \tan^{-1} \left ( \frac{|f|}{ L^2} \right )\right ) \frac{\left | \left( \partial_x f \right )^2 - \left( \partial_t f \right )^2 \right |^2 }{|f|} \Bigg \}
%\end{split}
%\end{equation}

\section{Magnetic soliton solution}
\label{MagneticSolitonSection}
\setcounter{equation}{0}

Given the correct K\"{a}hler metric with the proper periodicity conditions, we should be able to solve for the magnetic soliton solution that corresponds to a D1-string connecting the two intersecting D-branes.  The monopole is a half-BPS solution, so we can determine it by requiring that half the supersymmetry transformations still vanish.  We need to choose a BPS state whose unbroken supersymmetry is contained within the manifest ${\cal N}=1$.  The ${\cal N}=1$ supersymmetry algebra is the usual
$\{Q_\alpha, \bar Q_{\dot\beta}\} = 2 \sigma^\mu_{\alpha\dot\beta} P_\mu$, and so potential BPS states will have charges $T$-dual to $P_{2,3}$.  Thus we take the D3-branes to be separated in the 2-direction, implying expectation value for $A_{V2} - A_{V'2}$.  This is $SO(4)$-equivalent to the 8-direction depicted in Table~1, but has a different orientation with respect to the manifest ${\cal N}=1$.

Enforcing this, and demanding a static solution gives the following first order differential equations in $x^1$ for the monopole solution:
\be
\begin{split} 
&  \partial_1 Q_1 = \frac{1}{\sqrt{2}}  \partial_{z_1} A_{V2}  \,,
\\ & \partial_1 Q_2   = i \bar{\partial}_{z_1} \bar{Q}_3 \,,
\\ &  \partial_1 Q_3  =  -i  \bar{\partial}_{z_1} \bar{Q}_2 - \frac{1}{\sqrt{2}} \bar{f} \delta^2(x_4,x_5  ) \,,
\\ & \partial_1 A_{V2} = - \frac{\sqrt{2}}{2} \left ( \bar{\partial}_{z_1} Q_1 + \partial_{z_1} \bar{Q}_1 \right ) + \frac{1}{2}\mathcal{P} \delta^2(x_4,x_5  ) \,,
\\ & \partial_1 Z^\alpha  =  \frac{1}{\sqrt{2}} g^{\alpha \dot{\beta}} \partial_{\dot{\beta}} \bar{f} (\bar{Q}_3(0,0) - \bar S_3(0,0))  +  \frac{i}{2}  k^\alpha (A_{V2}(0,0) - A_{V'2}(0,0) ) \,,
\end{split}
\ee
with $A_{V0}=A_{V1}=A_{V3}=0$; similar equations hold for the $3'$-$3'$ fields.
These field equations contain $\ln(0)$'s from the D3 fields evaluated at the defect.  As usual, these are canceled by renormalization.  For example, $ A_{V2}(0,0) \sim {\cal P}\ln(0)$, which cancels against the field renormalization~(\ref{fieldren}).  %{{\color{red} Not sure about the $Q_3(0,0)$ term, I can massage it into almost the same form.}

To get some understanding of these equations, let us change the theory by taking the coefficients of the 3-3 and 3$'$-3$'$ actions to be large while holding fixed the 3-3$'$ action.  The 3+1 fields are now sourceless and we can set $A_{V2} = v$ with the others vanishing.  This leads to the simple differential equations
\be
\partial_1 B = \frac{v}{2} B \,\quad \partial_1 C = -\frac{v}{2} C \,. \label{bcsol}
\ee
The solutions are exponentials, which appear to blow up in one direction or the other.  To see that the solution is well-behaved, we must take into account the periodicity of $(B,C)$.  On the covering space, we have smooth coordinates $(B_n, C_n)$ in the neighborhood of the pole of the Gibbons-Hawking potential at $z = 2\pi n$.  The identification is $B_n = B_{n+1}/(\beta B_{n+1}C_{n+1}) = 1/\beta C_{n+1}$.  The solution $B_n = e^{vx^1/2}$ thus connects smoothly onto $C_{n+1} = e^{-vx^1/2}/\beta $ at large positive $x^1$.

We can also write this in terms of the $W,U$ coordinates.  The BPS equation is
\be
 \partial_1 Z^\alpha  =    \frac{iv}{2} k^\alpha \,, \quad k^W = -2i \,,\quad k^U = 0 \,.
\ee
 The solution is
\be
U(x) = U_0 \,,\quad W = vx^1 \,.
\ee
Noting the periodicity~(\ref{wperiod}) of $W$, this describes an infinite chain of solitons, spaced in $x^1$ by 
$ c\pi + \ln ({4\pi^2}/{|U_0^2|})$.  In the limit $U_0 \to 0$ the spacing becomes infinite, and we obtain a single D1. % {\color{ red} This behavior is a bit odd, why should the spacing be correlated with $U_0$?  I am not sure what coordinates to use to best describe the soliton, $U_0 \to 0$ is singular in some ways.} 

Let us also evaluate the solition mass in this model.  Defining $\partial_w = \partial_1 - i\partial_2$, the relevant term in the energy is proportional to
\begin{align}
&\int dx\,g_{\alpha\dot\beta} \left\{  (\partial_w Z^\alpha + i A_w \delta_\Lambda Z^\alpha)(\partial_{\bar w} Z^{\dot \beta} 
- i A_{\bar w} \delta_{\bar \Lambda} Z^{\dot\beta}) +
 (w \leftrightarrow \bar w) \right\} \nonumber\\
 =& \int dx\,2g_{\alpha\dot\beta} (\partial_w Z^\alpha + i A_w \delta_\Lambda Z^\alpha)(\partial_{\bar w} Z^{\dot \beta} 
- i A_{\bar w} \delta_{\bar \Lambda} Z^{\dot\beta}) + 2 v \partial_1 {\cal P} \,.
\end{align}
Thus the energy of a BPS state, for which the first term vanishes, is proportional to the magnetic charge.  

We have not found a simple treatment of the actual case of dynamical D3-brane fields.  However, by continuity the BPS soliton should continue to exist as the D3-brane $g_{\rm YM}$ is increased.  

There will also be dyonic solutions.  The soliton solution~(\ref{bcsol}) spontaneously breaks the $U(1)$ gauge symmetry, so constant $U(1)$ gauge transformations generate new solutions parameterized by a phase.  Quantizing this collective coordinate leads to the dyon spectrum.  While we have focused on the weak/strong element of the $S$-duality group, $2\pi$ shifts of the gauge theory $\theta$-parameter will generate the rest of the $SL(2,Z)$ action on the dyonic spectrum.

Dorigoni and Tong~\cite{Dorigoni:2014yfa} have noted that  the BPS equations do not have solutions corresponding to magnetic charge greater than 1, even though it might seem that one could form BPS states of multiple D1-branes connecting the D3 and D3'.  They give an elegant explanation for this: there is a repulsive force between D1-branes due to string disk instantons, bounded by a pair of D1's, the D3, and the D3'.  This is dual to the gauge instanton-induced superpotential of Ref.~\cite{Affleck:1982as}.

%We expect the solution to these differential equations to take the following form.  $f(BC)=BC$ should vanish everywhere, and thus so should $Q_3$ and $Q_2$.  This solves the second and third equations trivially.  In the brane picture, the corresponds the the branes not bending along each other.  We can integrate the last equation to get
%\be
%B = B_0 e^{xA_{V2}(0,0)/2} \quad \quad \quad C = C_0 e^{-xA_{V2}(0,0)/2}
%\ee 
%If the gauge field is weakly coupled to the intersection fields corrections to $A_{V2}$ from the first and fourth equations should be small, and to leading order we can take $B$ and $C$ to behave as exponentials.  $B$ or $C$ nonzero is only a good solution for $x<0$ and $x>0$ respectively.  To prevent the exponential blow up and maintain $BC=0$, take $B_0 =0$ for $x>0$ and $C_0 =0$ for $x<0$.  The inability for a single solution to cover the whole space is likely a coordinate issue.  Presumably, were we able to explicitly preform the transform to better coordinates such as the ones in Section \ref{TaubNutSection}, this problem would be resolved.  {\color{red} Is there anything to say about asymptotic values of $\mathcal{P}$?  We expect them to be different for $x\to\pm\infty$, but I'm not sure if we can say anything more explicit about its value in the limits.} 
\section{Discussion}
\label{DiscussionSection}

Our study of the D3-D3 intersection has led to a number of surprises.  The first is the magnetic couplings of the hypermultiplets.  In one sense this has a simple origin.  The hypermultiplet bosons are spinors in the ND directions, and so it is not surprising that they have magnetic dipole couplings.  In a sense, a line of dipoles is like a separated monopole-antimonopole pair.  (This is similar to the kink construction of fermions in bosonization, in that $\partial_\pm\phi$ produce dipoles of fermion number.)  The dipole density is $\cal P$, and the monopole density is ${\cal P}'$.  This does not seem to be a useful way to make magnetic monopoles in nature, but here when the dipole density $\cal P$ reaches a multiple of $8\pi$, the line disappears and we are left with unconfined monopoles.

The essential role of the nonlinear hypermultiplet kinetic term and the periodicity of the hypermultiplet moduli space is the second interesting feature.  Again, this is relevant only for intersections with two ND and two DN directions.  With three plus one the moduli space is $R^3$, and with four plus zero it is $R^4$ (in noncompact spaces).  This is for the Abelian case; we have not considered intersections of stacks of branes.

The third interesting feature is the need for classical renormalization.  Our results highlight the importance of this phenomenon~\cite{Goldberger:2001tn}, which may be useful for understanding other brane intersections, and brane self-interactions.

The final surprising feature is the inconsistency of the intersection theory as a field theory.  Note that the metric on moduli space is infinite, so the usual logarithmic exploration of the space of vacua will be absent.  Nevertheless, any attempt to define this theory via a path integral will run into the regions of negative metric.

Our consideration of this system was originally motivated by the construction of top-down AdS/CM models.  The magnetic source terms appeared implicitly in Refs.~\cite{Constable:2002xt,Jensen:2011su}, but the present interpretation was not noted.  It seems unlikely that these terms have relevance for condensed matter duals; rather, they exemplify some of the rigidity of top-down constructions.

Another motivation for this work was the exploration of simple models of electric-magnetic duality.  The lack of a UV completion in field theory complicates this question, but there may still be interesting questions in the effective field theory.

\section*{Acknowledgements}
We thank Ahmed Almheiri, Johanna Erdmenger, Aki Hashimoto, Andreas Karch, Ben Michel, Greg Moore, Hirosi Ooguri, David Tong, and Edward Witten for discussions and communications. The work of EM was supported in part by NSF grants PHY07-57035 and PHY13-16748.  The work of JP was supported in part by NSF grants PHY11-25915 (academic year) and PHY07-57035 and PHY13-16748 (summer).
The work of SS  was supported in part by U.S. DOE grant DE-FG02-00ER41132 and by a KITP Graduate Fellowship supported by NSF PHY11-25915.

\end{document}